\documentclass[12pt]{article}
\usepackage{amsmath}
\usepackage{amssymb}
\usepackage{epsfig}
\usepackage{euscript}
\usepackage{fancybox}
\usepackage{color}

\def\mathswitchr#1{\relax\ifmmode{\mathrm{#1}}\else$\mathrm{#1}$\fi}

\newcommand {\pslash}{\hbox{$\not\hbox{\kern-2.3pt $p$}$}}

\usepackage{cite}

\usepackage{epic}

\def\alf1{ {\alpha\over\pi} }

\begin{document}
\begin{titlepage}
\begin{flushright}
{\bf BU-HEPP-07-04 }\\
{\bf May, 2007}\\
\end{flushright}
 
\begin{center}
{\Large Quark masses and resummation in precision QCD theory$^{\dagger}$
}
\end{center}

\vspace{2mm}
\begin{center}
{\bf   B.F.L. Ward}\\
\vspace{2mm}
{\em Department of Physics,\\
 Baylor University, Waco, Texas, 76798-7316, USA}\\
\end{center}

\vspace{5mm}
\begin{center}
{\bf   Abstract}
\end{center}
It is shown that amplitude-based, exact
resummation tames the un-canceled IR divergences
at ${\cal O}(\alpha_s^2)$ in initial state radiation in QCD with 
massive quarks. Implications
for precision predictions for LHC physics are discussed.
 
\par 
\vspace{10mm}
\vspace{10mm}
\renewcommand{\baselinestretch}{0.1}
\footnoterule
\noindent
{\footnotesize
\begin{itemize}
\item[${\dagger}$]
Work partly supported by US DOE grant DE-FG02-05ER41399,
by the Polish Government grant No. 620/E-77/6.PR UE/DIE 188/2005-2008 and
by NATO grant PST.CLG.980342.
\end{itemize}
}

\end{titlepage}

\def\Kmax{K_{\rm max}}\def\ieps{{i\epsilon}}\def\rQCD{{\rm QCD}}
\renewcommand{\theequation}{\arabic{equation}}
\font\fortssbx=cmssbx10 scaled \magstep2
\renewcommand\thepage{}
\parskip.1truein\parindent=20pt\pagenumbering{arabic}\par
The era of precision QCD at the LHC, by which we mean 1\% or 
better precision tags on the theoretical predictions, presents us with the 
extremely challenging task of proving that a given theoretical precision tag
does in fact hold to that level. This means that all aspects 
of the standard formula for hadron-hadron scattering in perturbative 
QCD have to be examined for possible sources of uncertainty in the 
physical and technical precision
components of any quoted total theoretical precision tag.
In this connection, we note the standard practice of treating all
quarks in the initial state as massless. It would be desirable to put
an explicit error tag on this assumption by doing the respective
calculations with the respective quark masses at their known~\cite{qmass}
values and comparing the attendant predictions with their massless limits.
This direct approach is however currently blocked by the pioneering
results in Refs.~\cite{chris1,cat1}, wherein it has been established that
there is a lack of Bloch-Nordsieck cancellation at ${\cal O}(\alpha_s^2)$
in the initial state radiation in massive QCD. Hence, even the $b$ quark
has to have zero mass in the initial state radiative corrections
when one works to ${\cal O}(\alpha_s^n)$ with $n\ge 2$.\par 

In what follows, we re-visit the results in Refs.~\cite{chris1} from 
the standpoint of recent progress~\cite{qcdexp,irdglap} in the resummation 
of large IR effects in the QCD 
perturbation theory, where we will focus on exact resummation methods
\footnote{We do not employ here the resummation algebras that focus on
the boundary of phase space~\cite{sterman,cattrent}, as these 
engender approximations
which require arguments that are not opportune for our purposes.}
with an eye toward rigorous control on any theoretical precision error budget
that we may ultimately want to advocate. In this context, let us recall
already the master formula that we have derived in Refs.~\cite{qcdexp}:
using a $2\rightarrow 2+X$ hard process with multiple gluon $(G)$ emission, 
$q(p_1)+q'(q_1)\rightarrow q''(p_2)+q'''(q_2)+n(G)+X(p_X)$ 
in an obvious 4-momentum
assignment notation, we have the differential cross section
\begin{equation}
\begin{split}
d\hat\sigma_{\rm exp}
         &=e^{\rm SUM_{IR}(QCD)}\sum_{n=0}^\infty\frac{1}{n!}\int\prod_{j=1}^n{d^3
k_j\over k_j}\int{d^4y\over(2\pi)^4}e^{iy\cdot(p_1+q_1-p_2-q_2-p_X-\sum k_j)+
D_\rQCD}\\
&*\tilde{\bar\beta}_n(k_1,\ldots,k_n){d^3p_2\over p_2^{\,0}}{d^3q_2\over
q_2^{\,0}}{d^3p_X\over p_X^{\,0}}
\end{split}
\label{subp15}
\end{equation}
where the hard gluon residuals 
$\tilde{\bar\beta}_n(k_1,\ldots,k_n)$
and the infrared functions ${\rm SUM_{IR}(QCD)},~D_\rQCD$ are defined in Ref.~\cite{qcdexp} 
and we stress that the $\tilde{\bar\beta}_n(k_1,\ldots,k_n)$
are free of all infrared divergences to all 
orders in $\alpha_s(Q)$. (See especially Ref.~\cite{irdglap} 
for explicit application
of (\ref{subp15}) to a real bremsstrahlung process.) 
Note that the hard gluon residuals $\tilde{\bar\beta}_n(k_1,\ldots,k_n)$
have the structure~\cite{qcdexp} \[\tilde{\bar\beta}_n(k_1,\ldots,k_n)=[tr\mathfrak{h}{\cal M}^{(n)\dagger}{\cal M}^{(n)}]|_{IR-subtracted}\] where the IR-subtraction is as given
in Ref.~\cite{qcdexp} and $\mathfrak{h}$ is the initial-state color-spin density matrix so that the 
full quantum mechanical color
effects are included in (\ref{subp15}). 
For our initial state radiation (ISR) analysis, we take q and q' to be massive
quarks of mass $m_q$, we take $X$ to be a (QCD singlet) electroweak gauge boson to match the problem studied in Ref.~\cite{chris1} and we 
only compute ISR radiative effects in QCD. Let us then recall the pioneering
result in Refs.~\cite{chris1}: working in the eikonal approximation
and discussing the part of the cross section proportional to 
the color structure (here, $H$ corresponds to the attendant hard sub-process) 
\begin{equation}
F_1=C_2(G)H^{\alpha\beta}_{ab}(T_i)^{\beta\alpha}(T_i)_{ba}
\end{equation}
for the process
$q^\alpha q^a \rightarrow V^{(*)}+X'$
where $V^{(*)}$ is our (off-shell) electroweak gauge boson and $\alpha,~a$
are the colors of the quarks, the authors in 
Refs.~\cite{chris1} find the IR divergent result
\begin{equation}
\text{flux}~\frac{d\sigma}{d^3 Q}=\frac{-g^4 \bar{H}}{(d-4)32\pi^2}\left(\frac{1-\beta}{\beta}\right)\left(\frac{1}{\beta}\ln(\frac{1+\beta}{1-\beta})-2\right)
\label{noBNcancl1}
\end{equation},
where $g$ is the QCD coupling constant, $\bar{H}$ is the attendant hard
sub-process factor dressed as in the color structure $F_1$, $\beta$ is the
velocity of one quark in the rest frame of the other, and $d$ is the 
dimension of space-time with $d>4$ to regulate the uncanceled IR
divergence. $Q^2$ is the invariant mass of the $V^{*}$. This divergence
is clearly non-Abelian in character as it vanishes for $C_2(G)=0$, where
we define the gluon and quark representations' quadratic Casimir invariants
respectively as usual:
\begin{equation}
\begin{split}
f_{ijk}f_{ijl}&=C_2(G)\delta_{kl}\cr
(T_iT_i)_{ab}&=C_2(F)I_{ab}\equiv C_FI_{ab},
\end{split}
\label{eq-clr1}
\end{equation}
where $f_{ijk}$ are the group structure constants.
The result (\ref{noBNcancl1}) shows a clear lack of Bloch-Nordsieck 
cancellation at ${\cal O}(\alpha_s^2)$ and the standard approach is
to set $m_q=0$ so that this uncanceled IR divergence vanishes as $\beta\rightarrow 1$ as one can see from (\ref{noBNcancl1}).\par
We point-out that the authors in Ref.~\cite{cat1} have analyzed 
the problem studied in Refs.~\cite{chris1} from a coherent-state Hamiltonian
approach and have corroborated the result (\ref{noBNcancl1}) with the added
understanding that, in the coherent-state approach, the single pole
divergence is converted into an unfactorizable, unspecified dependence 
of the respective collinear singularities on the scale
separating the attendant observable and un-observable gluon degrees of freedom
inherent therein. This is again unacceptable and forces the 
use of $m_q=0$ for initial ISR for calculations at 
${\cal O}(\alpha_s^n),~n\ge 2$.\par 
Here we propose an alternative approach. We look into the 
systematics of the analysis of 
the first paper in Refs.~\cite{chris1}. We see that one can represent the
RHS of (\ref{noBNcancl1}) as the left-over real IR divergence 
which is uncanceled by the virtual IR divergence. In the language
of the diagrams analyzed by Mueller's theorem~\cite{mueller} 
in the aforementioned paper, the RHS of (\ref{noBNcancl1}) can be identified
with a fraction $F_{nbn}$ of the contribution of the
real emission from the contribution of the diagrams equivalent to 
the diagram contribution (q-o) in Fig. 6 in the first paper in
Ref.~\cite{chris1}, which we
reproduce here for definiteness in Fig.~1.
\begin{figure}
\begin{center}
\epsfig{file=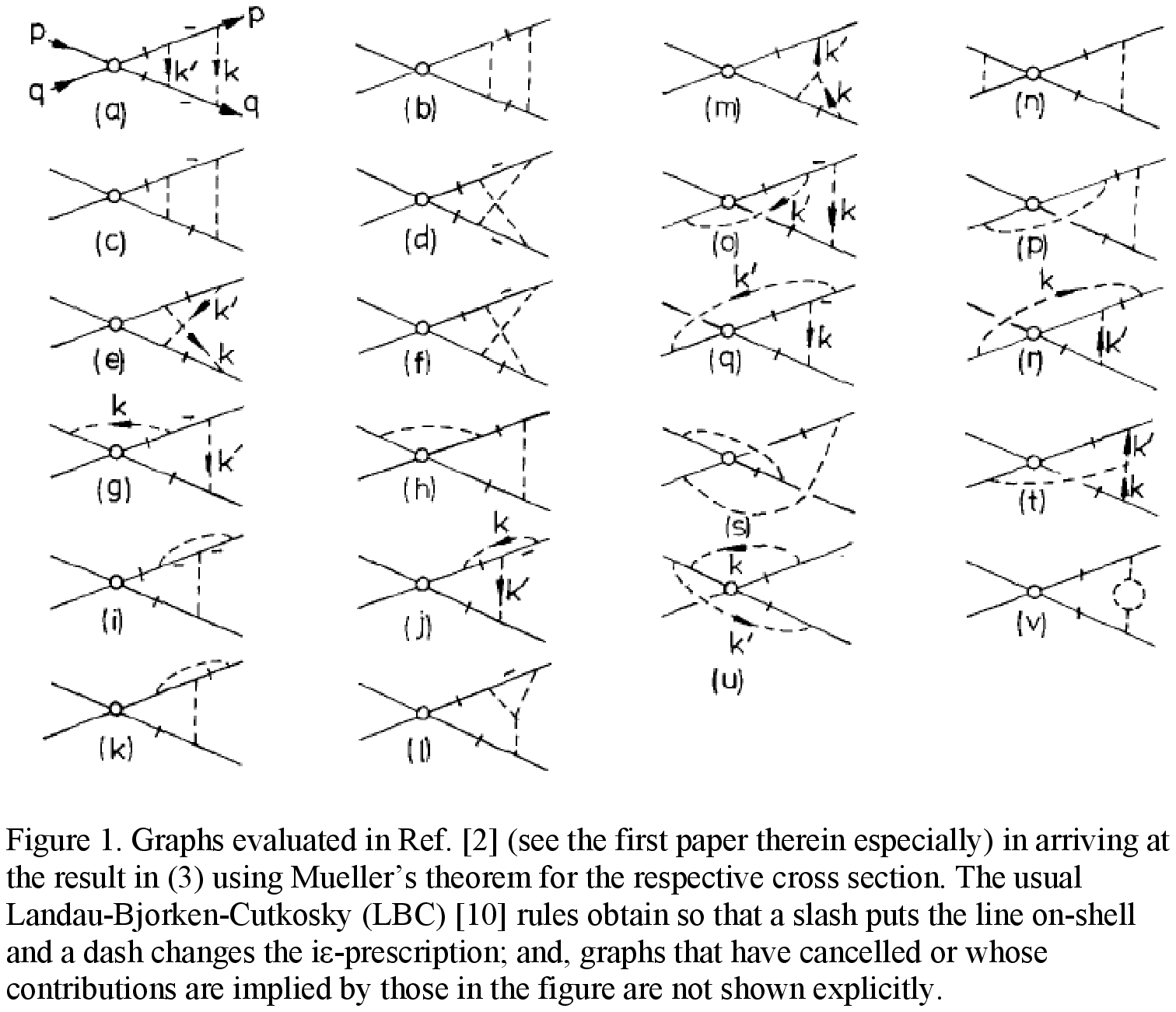,width=15cm}
\end{center}
\end{figure}
\noindent
To see this, let us recall the result of this last paper for the
(q-o) diagrams (see Fig.~1)
contribution to the 
differential cross section, removing
the kinematic (note in this language the hard scattering factor
is kinematic to the soft interactions under study here) 
and color factors: from the 5th equation on page
11 of the paper, we have the result, from the equivalence of
diagrams c and q and the equivalence of diagrams o and f in
Fig. 6 of the paper (Fig.~1 here),
\begin{equation}
A_{q-o}=\frac{1}{\beta^2}\int\frac{d^3kd^3k'2k_z}{(k_z+{k'}_z+i\epsilon)(\beta^2k_z^2-{\bf k}^2)(\beta^2k_z^2-{\bf k'}^2+i\epsilon)(k_z^2+\epsilon^2)},
\label{eqn:q-o}
\end{equation}
where we denote the 3-momentum by boldface letters, so that $\vec{k}={\bf k}$,
and where the eikonal limit has been used in (\ref{eqn:q-o}) as it was
in respective paper in
Ref.~\cite{chris1}. Because of this approximation, there is a spurious  
UV divergence in (\ref{eqn:q-o}), which does not affect the IR regime. The
authors in Ref.~\cite{chris1} therefore regulate this UV divergence with
the factor $e^{-{\bf k}^2/\Lambda^2}$ for each would-be 
3-space integral and then use
dimensional methods~\cite{tHvelt} to isolate the IR divergence of interest;  
one obtains in this way the UV regulated result from (\ref{eqn:q-o})
\begin{equation}
A_{q-o}|_{UV-reg}=\frac{4\pi^{n+1}(\Lambda^2)^{n-3}}{\beta^2}\left\{\frac{1}{(n-3)^2}+\frac{1}{2(n-3)}\ln\left(\frac{1+\beta}{1-\beta}\right)\right\},
\label{eqn:q-o-1}
\end{equation}
where the UV cut-off $\Lambda$ is large compared to the soft scales in the
problem, and here we have $d=n+1$ to make contact with (\ref{noBNcancl1}).
When one adds the remaining contributions associated to the remaining graphs in
Fig.~1,
one sees from comparing (\ref{noBNcancl1}) and
(\ref{eqn:q-o-1}) that the double pole term in (\ref{eqn:q-o-1})
is canceled and that the fraction
\begin{equation}
F_{nbn}= \frac{(1-\beta)(\ln\left(\frac{1+\beta}{1-\beta}\right)-2\beta)}{\ln\left(\frac{1+\beta}{1-\beta}\right)} 
\end{equation}
of the single pole term is left over as the uncanceled IR divergence.\par

The classic Landau-Bjorken-Cutkosky (LBC) 
analysis then allows us to determine the
relationship between the real emission in the (q-o) diagram contribution
and the single pole term on the RHS of (\ref{eqn:q-o-1}). Specifically,
upon doing the integral on the RHS of (\ref{eqn:q-o}) over ${k'}_z$, there are
are two poles in the respective complex plane below the real axis, one
at $-k_z-i\epsilon$ and one at $-\sqrt{\beta^2k_z^2-{{k'}_\perp}^2+i\epsilon}$,
where here the energy of the $k'\text{-gluon}$ is just $-\beta k_z$ by the
LBC rules in this eikonal exercise. The contribution of the
former pole does not result in on-shell $k'$ gluons
. The LBC rules tell us
that the regime ${\mathfrak R}=\{0\le {{k'}_\perp}^2\le \beta^2k_z^2\}$
represents the regime wherein the $k'\text{-gluon}$ is actually on-shell
here. Focusing on this regime, we see that we have the contribution
\begin{equation}
\begin{split}
A_{q-o}|_{\mathfrak R}&=\Re \frac{1}{\beta^2}\int d^3k\int_0^{\beta^2k_z^2}\pi d({{k'}_\perp}^2)\frac{-2\pi i}{-(-2)\sqrt{\beta^2k_z^2-{{k'}_\perp}^2}}\frac{1}{k_z-\sqrt{\beta^2k_z^2-{{k'}_\perp}^2}+i\epsilon}\\
&\qquad \frac{1}{\beta^2k_z^2-{\bf k}^2}\frac{2k_z}{k_z^2+\epsilon^2},
\end{split}
\label{eqn:q-o-2}
\end{equation}  
where we have written the 2-space 
integration measure as $\pi d({{k'}_\perp}^2)$
by doing the respective angular integral. The integration over the 
latter measure can then be re-written, using the fact that
we only need the real part, 
\begin{equation}
\begin{split}
A_{q-o}|_{\mathfrak R}&=\Re \frac{-\pi i}{\beta^2}\int d^3k\int_0^{\beta^2k_z^2}\pi d({{k'}_\perp}^2)\frac{1}{\sqrt{\beta^2k_z^2-{{k'}_\perp}^2}}\frac{k_z+i\epsilon+\sqrt{\beta^2k_z^2-{{k'}_\perp}^2}}{(k_z+i\epsilon)^2-(\beta^2k_z^2-{{k'}_\perp}^2)}\\
&\qquad \qquad\qquad\frac{1}{\beta^2k_z^2-{\bf k}^2}\frac{2k_z}{k_z^2+\epsilon^2}\\
         &= \Re \frac{-\pi i}{\beta^2}\int d^3k\int_0^{\beta^2k_z^2}\pi d({{k'}_\perp}^2)\frac{1}{\sqrt{\beta^2k_z^2-{{k'}_\perp}^2}}\frac{k_z+i\epsilon}{(k_z+i\epsilon)^2-(\beta^2k_z^2-{{k'}_\perp}^2)}\\
&\qquad \qquad\qquad \frac{1}{\beta^2k_z^2-{\bf k}^2}\frac{2k_z}{k_z^2+\epsilon^2}\\
        &=\Re \frac{-\pi i}{\beta^2}\int d^3k\int_0^{\beta^2k_z^2}\pi d({{k'}_\perp}^2)\frac{1}{\sqrt{\beta^2k_z^2-{{k'}_\perp}^2}}\\
&\qquad \frac{1}{2}\left(\frac{1}{k_z+i\epsilon-\sqrt{\beta^2k_z^2-{{k'}_\perp}^2}}+\frac{1}{k_z+i\epsilon+\sqrt{\beta^2k_z^2-{{k'}_\perp}^2}}\right)\frac{1}{\beta^2k_z^2-{\bf k}^2}\frac{2k_z}{k_z^2+\epsilon^2}\\
       &=\Re \frac{-i\pi^2}{\beta^2}\int d^3k\left(-\ln(k_z+i\epsilon-\beta|k_z|)+\ln(k_z+i\epsilon+\beta|k_z|)\right)\frac{1}{\beta^2k_z^2-{\bf k}^2}\frac{2k_z}{k_z^2+\epsilon^2},
\end{split}
\label{eqn:q-o-3}
\end{equation}
where we again emphasize that the on-shell regime actually has $k'_0=-\beta k_z<0$ so that the real radiative contribution, by the standard LBC methods,
has $k_z>0$. If we integrate over the region $k_z>\sqrt{\epsilon}$, 
it is clear that the RHS of the
last equation has no real part as $\epsilon\rightarrow 0$. Thus, 
the real emission part of (\ref{eqn:q-o-3}) must arise from the regime
$0\le k_z\le \sqrt{\epsilon}$. We treat the branch cuts for the logs by
joining them between $k_{z1}=-i\epsilon/(1-\beta)$ and $k_{z2}=-i\epsilon/(1+\beta)$ and then we close the contour below the real axis as shown in
\begin{figure}
\begin{center}
\epsfig{file=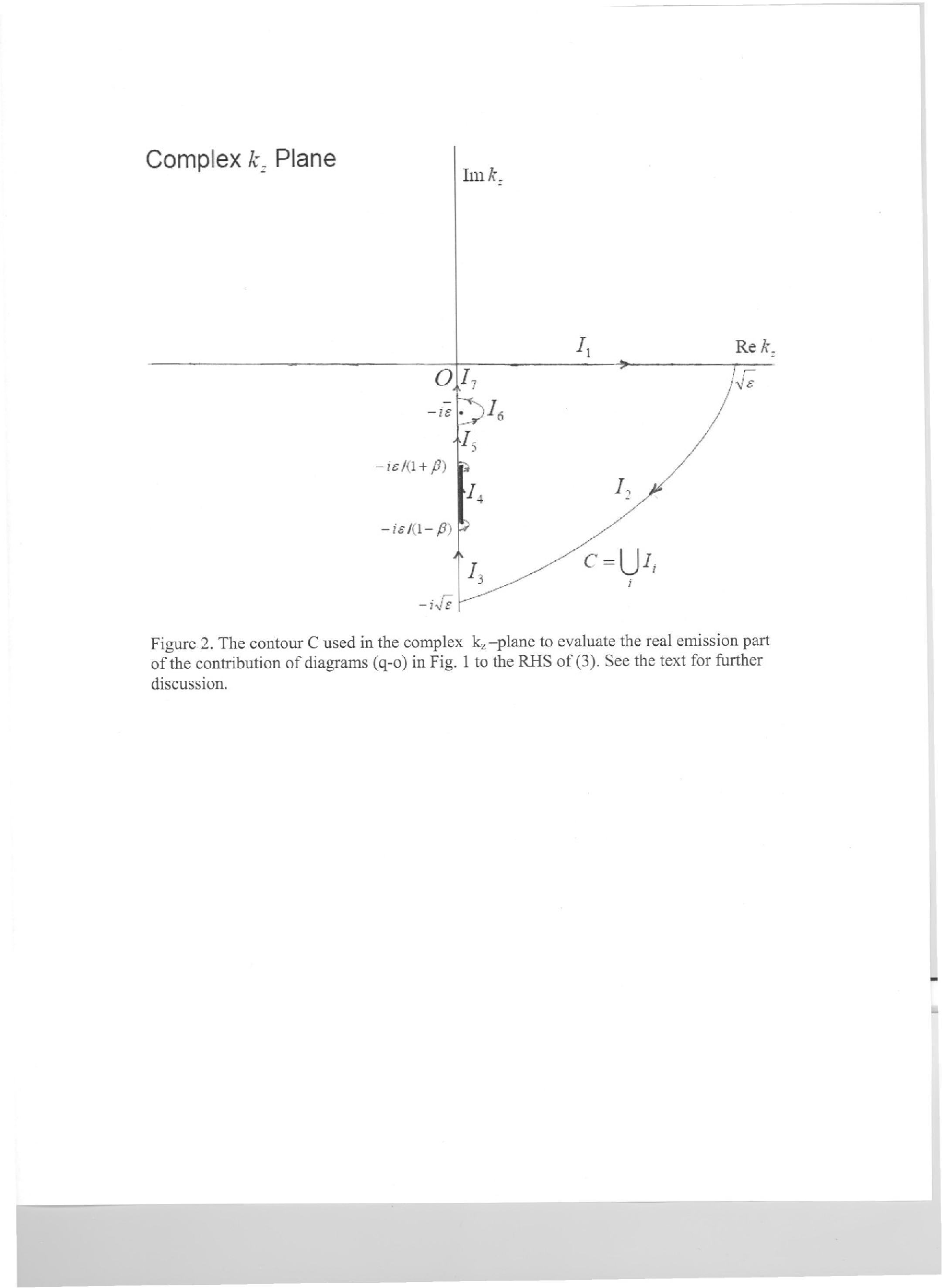,width=15cm}
\end{center}
\end{figure}
\noindent
Fig.~2
to get the result, by Cauchy's theorem,
\begin{equation}
\oint_Cdk_z\left(-\ln(k_z+i\epsilon-\beta k_z)+\ln(k_z+i\epsilon+\beta k_z)\right)\frac{1}{\beta^2k_z^2-{\bf k}^2}\frac{2k_z}{k_z^2+\bar{\epsilon}^2}=0,
\label{eqn:q-o-4}
\end{equation}
where we use the intrinsic freedom in the Feynman $i\epsilon$-prescription
to take each such infinitesimal parameter independently to 0 from above
and the curve $C$ is given in Fig.~2.
We take here $k_\perp>\sqrt{\epsilon}$.\footnote{We use standard 
Lebesgue integration theory to conclude that the order of integration
does not matter so that fixing $k_\perp$ and integrating over $k_z$ first, 
which means that the limit $\epsilon \rightarrow 0$ will always give us $k_\perp> \sqrt{\epsilon}$, followed by integration over the full range of $k_\perp$, when all integrals are finite 
by regularization where necessary,
can not affect the final result. Alternatively, the reader can check that
with the regularization we use, if one does the attendant 
integral over $0\le k_\perp\le\sqrt{\epsilon}$, the respective result will vanish for $\epsilon\rightarrow 0$.} If we denote the integrals
over the $i-th$ part of $C$ by $I_i,~i=1,\cdots,7$, where the labels for these
parts are defined in Fig.~2,
then one can readily see
that (for example we may set $\bar{\epsilon}=\epsilon^{\frac{3}{2}}$) 
we have
\begin{equation}
\begin{split}
I_1&=\int_0^{\sqrt{\epsilon}} dk_z\left(-\ln(k_z+i\epsilon-\beta|k_z|)+\ln(k_z+i\epsilon+\beta|k_z|)\right)\frac{1}{\beta^2k_z^2-{\bf k}^2}\frac{2k_z}{k_z^2+\bar{\epsilon}^2}\\
&=-\sum_{i=2}^{7}I_i.
\end{split}
\label{eqn:q-o-5}
\end{equation}
We now treat the integrals $I_i,~i=2,\cdots,7$ in turn.\par
For $I_2$, use the change of variable $k_z=\sqrt{\epsilon}e^{i\theta}$, for
$0\ge\theta\ge -\frac{\pi}{2}$. Then, we get
\begin{equation}
\begin{split}
I_2&=\int_0^{-\frac{\pi}{2}} id\theta k_z\left(-\ln(k_z+i\epsilon-\beta k_z)+\ln(k_z+i\epsilon+\beta k_z)\right)\frac{1}{\beta^2k_z^2-{\bf k}^2}\frac{2k_z}{k_z^2+\bar{\epsilon}^2}\\
&=2\int_0^{-\frac{\pi}{2}} id\theta \left(-\ln(1-\beta)+\ln(1+\beta)\right)\frac{1}{-{\bf k_\perp}^2}\\
&=-i\pi\ln\left(\frac{1+\beta}{1-\beta}\right)\frac{1}{(-{\bf k_\perp}^2)}.
\end{split}
\label{eqn:q-o-6}
\end{equation}
\par
For $I_3$ it is enough to use the change of variable $k_z=-iy$ to see that it
is pure real so that it will not contribute to the imaginary part
of $I_1$ via (\ref{eqn:q-o-4}) and only this part of $I_1$ is needed
in extracting the real emission part of the RHS of (\ref{eqn:q-o-3}).\par
For $I_4$ we see from passing around the lower branch point
in Fig.~2
that the respective imaginary contribution
is just
\begin{equation}
\begin{split}
i\Im I_4&=\int_{\frac{-i\epsilon}{1-\beta}}^{\frac{-i\epsilon}{1+\beta}} dk_z\left(-\pi i\right)\frac{1}{(-{\bf k_\perp}^2)}\frac{2k_z}{k_z^2+\bar{\epsilon}^2}\\
&=2\pi i\ln\left(\frac{1+\beta}{1-\beta}\right)\frac{1}{(-{\bf k_\perp}^2)}.
\end{split}
\label{eqn:q-o-7}
\end{equation}
\par
For $I_5$, we see by the change of variable $k_z=-iy$ that it is pure real
and does not contribute to the imaginary part of $I_1$ via (\ref{eqn:q-o-4}).
\par
For $I_6$, we get the result
\begin{equation}
\begin{split}
I_6&= \pi i Res(-i\bar{\epsilon})\\
   &=0
\end{split}
\label{eqn:q-o-8}
\end{equation}  
since $\bar{\epsilon}/\epsilon\rightarrow 0$ when $\epsilon\rightarrow 0$.
\par
Finally, for $I_7$ the change of variable $k_z=-iy$ shows that
it too is pure real and does not contribute to the imaginary part
of $I_1$ via (\ref{eqn:q-o-4}).\par
The net result is that we arrive at
\begin{equation}
\begin{split}
i\Im I_1&= -\{2\pi i-\pi i\}\frac{-1}{{\bf k_\perp}^2}\ln\left(\frac{1+\beta}{1-\beta}\right) \\
   &= \frac{\pi i}{{\bf k_\perp}^2}\ln\left(\frac{1+\beta}{1-\beta}\right).
\end{split}
\label{eqn:q-o-9}
\end{equation} 
\par
When we introduce the RHS of (\ref{eqn:q-o-9}) into (\ref{eqn:q-o-3})
we get the result
\begin{equation}
A_{q-o}|_{{\mathfrak R},\text{real rad.}}=\frac{2\pi^3}{\beta^2}\left(\frac{1}{2}\ln\left(\frac{1+\beta}{1-\beta}\right)\right)\int \frac{d^2k_\perp}{{\bf k_\perp}^2},
\label{eqn:q-o-10}
\end{equation}
where we explicitly indicate that this is the real emission contribution
by the subscript $\text{real rad.}$. Using the UV regulator 
employed in the
first paper in Refs.~\cite{chris1}, we see 
that the integral over ${\bf k_\perp}$
in (\ref{eqn:q-o-10}) can be written as 
\begin{equation}
\begin{split}
{\cal I}_{\text{UV reg.}}&=\int \frac{d^2k_\perp e^{-{\bf k_\perp}^2/\Lambda^2}}{{\bf k_\perp}^2}\\
&=\int \frac{d^3k\delta(k_z)e^{-{\bf k}^2/\Lambda^2}}{{\bf k}^2}.
\end{split}
\end{equation}
We regulate the infrared divergence by analytic continuation to n dimensions
to get
\begin{equation}
\begin{split}
{\cal I}_{\text{UV reg.,IR reg.}}&=\int \frac{d^nk\delta(k_z)e^{-{\bf k}^2/\Lambda^2}}{{\bf k}^2}\\
&=\int_0^\infty d\rho\int d^nk\delta(k_z)e^{-{\bf k}^2/\Lambda^2-\rho{\bf k}^2}\\
&=\frac{2\pi^{\frac{(n-1)}{2}}}{n-3}(\Lambda^2)^{\frac{n-3}{2}}.
\end{split}
\label{eqn:q-o-11}
\end{equation}
\par
Introducing this last result into (\ref{eqn:q-o-10}), we get
\begin{equation}
A_{q-o}|_{{\mathfrak R},\text{real rad., UV reg.}}=\frac{4\pi^4(\pi\Lambda^2)^{\frac{n-3}{2}}}{\beta^2}\left(\frac{1}{2(n-3)}\ln\left(\frac{1+\beta}{1-\beta}\right)\right),
\label{eqn:q-o-12}
\end{equation}
which shows that the real emission part of $A_{q-o}$ saturates its 
single IR pole contribution.\par
Isolating the divergent single pole IR term in (\ref{eqn:q-o-12}) we may now
re-write the pioneering result of Refs.~\cite{chris1} as follows: the
uncanceled IR singular contribution to the respective differential 
cross section is 
\begin{equation}
\text{flux}~\frac{d\sigma}{d^3 Q}=\frac{-g^4 \bar{H}}{64\pi^6}F_{nbn}A_{q-o}|_{{\mathfrak R},\text{real rad., IR pole part}},
\label{noBNcancl2}
\end{equation}
where from (\ref{eqn:q-o-12}) we have
\begin{equation}
A_{q-o}|_{{\mathfrak R},\text{real rad., IR pole part}}=\frac{4\pi^4}{\beta^2}\left(\frac{1}{2(n-3)}\ln\left(\frac{1+\beta}{1-\beta}\right)\right).
\label{eqn:q-o-13}
\end{equation}
We note that the result in (\ref{eqn:q-o-12}) agrees with the single pole
term in (\ref{eqn:q-o-1}) and with (\ref{eqn:q-o-13}) up to finite terms.
\par
From the result (\ref{noBNcancl2}) we can now see how the theory of exact,
amplitude resummation may impact the conclusions of Refs.~\cite{chris1}.
We apply the formula in (\ref{subp15}) to the real emission process in
$A_{q-o}|_{\mathfrak R}$, following for example the steps given in
in Ref.~\cite{irdglap}. We stress that we apply the resummation
only to that fraction, $F_{nbn}$, of the real emission 
that has the uncanceled IR singularity in (\ref{noBNcancl2}). The
remaining $1-F_{nbn}$ is not resummed because it is canceled by
the sum of the remaining contributions associated with the diagrams in
Fig.~1.
We get, in this way, the result
\begin{equation}
\begin{split}
F_{nbn}A_{q-o}|_{{\mathfrak R},\text{real rad., resummed}}&=F_{nbn}\Re \frac{-i\pi^2}{\beta^2}\int d^2k_\perp\int_0^{\sqrt{\epsilon}}dk_zF_{YFS}(\gamma_q)e^{\delta_q/2}(\beta k_z)^{\gamma_q}\\
&\qquad\left(-\ln(k_z+i\epsilon-\beta k_z)
+\ln(k_z+i\epsilon+\beta k_z)\right)\\
&\qquad\frac{1}{\beta^2k_z^2-{\bf k}^2}\frac{2k_z}{k_z^2+\epsilon^2},
\end{split}
\label{eqn:q-o-14}
\end{equation}
where we have defined the resummation functions, from Ref.~\cite{irdglap},
\begin{align}
\gamma_q &= 2C_F\frac{\alpha_s(Q^2)}{\pi}(\ln(s/m^2)-1)\\
\delta_q&=\frac{\gamma_q}{2}+\frac{2\alpha_sC_F}{\pi}(\frac{\pi^2}{3}-\frac{1}{2})
\label{dglap9}
\end{align}
and 
\begin{equation}
F_{YFS}(\gamma_q)=\frac{e^{-C_E\gamma_q}}{\Gamma(1+\gamma_q)}.
\label{dglap10}
\end{equation}
Here, $C_F$ is the quark representation 
quadratic Casimir invariant already defined in (\ref{eq-clr1}),
$s=(p_1+q_1)^2$ in our process in (\ref{subp15})
specialized to $X=V^{(*)}$ with $Q^2=p_X^2$, 
\[C_E=.5772\dots\] is Euler's constant
and $\Gamma(w)$ is Euler's gamma function. The function
$F_{YFS}(z)$ was already introduced by Yennie, Frautschi
and Suura~\cite{yfs} in their analysis of the IR behavior
of QED.
Using the substitution $k_z=\sqrt{\epsilon}\bar{k}_z$, we have 
\begin{equation}
\begin{split}
F_{nbn}A_{q-o}|_{{\mathfrak R},\text{real rad., resummed}}&=F_{nbn}\Re \frac{-i\pi^2\epsilon^{\frac{\gamma_q}{2}}}{\beta^2}\int d^2k_\perp\int_0^{1}d\bar{k}_zF_{YFS}(\gamma_q)e^{\delta_q/2}{(\beta\bar{k}_z)}^{\gamma_q}\\
&\qquad\left(-\ln(\bar{k}_z+i\sqrt{\epsilon}-\beta \bar{k}_z)+\ln(\bar{k}_z+i\sqrt{\epsilon}+\beta \bar{k}_z)\right)\\
&\qquad\frac{1}{-(1-\beta^2)\epsilon{\bar{k}_z}^2-{\bf k_\perp}^2}\frac{2\bar{k}_z}{{\bar{k}_z}^2+\epsilon}.
\end{split}
\label{eqn:q-o-15}
\end{equation}
We see that the RHS of this last equation vanishes as $\epsilon\rightarrow 0$,
removing the violation of Bloch-Nordsieck cancellation in (\ref{noBNcancl2}),
and, thereby, in (\ref{noBNcancl1})\footnote{Note that, by the mean value theorem, the RHS of (\ref{eqn:q-o-15}) is equal to $\epsilon^{\gamma_q/2}F_{YFS}(\gamma_q)e^{\delta_q/2}<{(\beta\bar{k}_z)}^{\gamma_q}>F_{nbn}A_{q-o}|_{{\mathfrak R},\text{real rad., IR pole part}}$, where $<A>$ denotes the respective mean value of $A$ defined with $d>4$;
thus, (\ref{eqn:q-o-15}) is still a higher twist effect with a coefficient
which vanishes as $\epsilon\rightarrow 0$.}. 
\par

We conclude that the result in Refs.~\cite{chris1} is obviated by amplitude
based exact resummation of the higher order corrections in QCD perturbation
theory. Only the infrared singular term from (\ref{eqn:q-o-12}) is
exponentiated, so that the finite non-zero terms in the cross section
are all treated on equal footing -- there is then no scheme dependence
introduced by our resummation.
The way is open to employ the current quark masses
in ISR phenomenology for the LHC. For the light quarks,
their main use will be as collinear/IR regulators, as the usual 
factorization
methods~\cite{qcdfactorzn} will generally replace them with the scale of such factorization;
for the b quark, we can not exclude at this time that
its mass may have some additional role in precision LHC theory.
Indeed, in addition to current algebra constraints, 
we know that from
the measured differences between the parton densities for s, c, and b
quarks in the proton that the ``heavy'' quark masses can not actually be zero.
We follow Ref.~\cite{collins}\footnote{In the proof of factorization presented
in Ref.~\cite{collins} for heavy quarks, there is an implicit use
of the cancellation of ISR infrared singularities; our results
remove any issues concerning this use.} in defining parton densities for heavy
quarks here. The issue then is the accuracy of the 
massless approximation in the ISR
in the context of precision LHC physics; for, already in QED, it is known
that the corresponding limit $m_e \downarrow 0$ in ISR and the condition
$m_e=0$ in ISR differ in ${\cal O}(\alpha/\pi)$. In QCD $\alpha_s/\pi\cong 3\%$
at TeV scales and this would be unacceptable if it would occur when the
precision tag is 1\%, as it will be at the LHC for some processes.\par 
We note here that there is  
considerable literature~\cite{w-k1,w-k2,w-k3,s-c-k,
krte-sch,thrn}
on the use of
quark masses in perturbative QCD phenomenology, especially for deep inelastic
scattering (DIS) processes. While in the original ACOT~\cite{w-k1} variable flavor number scheme
and in Ref.~\cite{krte-sch}, quark masses are retained in the
initial state analysis, in most cases,
following 
the S-ACOT~\cite{s-c-k} variable flavor number scheme and various extensions~\cite{w-k3,thrn}, the ISR is treated with zero quark mass in the
hard scattering coefficient with
possible use an appropriate rescaling variable $x(1+4m^2/Q^2)$~\cite{w-k2},
in standard DIS notation. 
These anaylses result 
in general in a better fit to the available
structure function data, although for Ref.~\cite{w-k3} the significance
of the attendant improved $\chi^2$ is within the range of uncertainty
of the respective fully massless result.
These efforts all speak to the need for proper treatment of quark mass
effects in precision high energy QCD phenomenology.\par  
We have discussed the theorem in Refs.~\cite{chris1} in which the
Drell-Yan process for quark-quark scattering is considered. However,
our solution for the lack of Bloch-Nordsieck cancellation only depended
on the external lines in the initial state, so it will carry-over to 
all such ISR configurations: exponentiation of real corrections
will render an extra factor of $k_0^{\gamma_q}$ in the respective integral
over phase-space to remove any end-point contributions which are
not already canceled by virtual corrections as required by the
Bloch-Nordsieck theorem.\par
Further implications of the results in 
this paper will appear elsewhere.~\cite{elswh}  
\section*{Acknowledgments}
We thank Profs.S. Jadach and S. Yost for useful discussions and 
Profs. L. Alvarez-Gaume and W. Hollik, respectively,
for the kind hospitality of the CERN TH Division and the 
Max-Planck-Institut, Munich, wherein
a part of this work was completed.

\newpage
\section*{Figure Captions}
\noindent
Figure 1. Graphs evaluated in Ref.~\cite{chris1}
(see the first paper therein, especially) 
in arriving at the result in(\ref{noBNcancl1})
using Mueller's theorem for the respective cross section. Here, the usual
Landau-Bjorken-Cutkosky(LBC)~\cite{blc} rules obtain 
so that a slash puts the line on-shell 
and a dash changes the $i\epsilon$ prescription; and graphs that 
have canceled or whose contributions are implied by those in  
the figure are not shown explicitly.
\par
\noindent 
Figure 2. The contour $C$ used in the complex $k_z$-plane to evaluate the
real emission part of the contribution of diagrams (q-o)
in Fig.~1
to the RHS of (\protect\ref{noBNcancl1}).
See the text for further discussion.\par
\end{document}